\documentclass[12pt]{article}

\usepackage[cp1251]{inputenc}
\usepackage{amssymb,amsmath}
\usepackage{hyperref}
\usepackage[mathscr]{eucal}
\usepackage{multicol}
\usepackage{graphicx}
\usepackage{amsfonts}
\usepackage{longtable}

\usepackage{epic}
\usepackage{eepic}
\usepackage{epsf}
\newcommand{\bU}{\bar{U}}
\newcommand{\hU}{\hat{U}}
\newcommand{\bV}{\bar{V}}
\newcommand{\hV}{\hat{V}}

\newsavebox{\risunok}

\newsavebox{\diag}

\begin{document}

\title{\vspace{3cm}Descent Relations and Oscillator Level Truncation Method}

\author{\vspace{1cm} \text{I.Ya.~Aref'eva${}^1$, R.~Gorbachev${}^{1}$,
P.B.~Medvedev${}^{2}$, D.V.~Rychkov${}^{3}$}\\
\\
\textit{\small ${}^1$ Steklov Mathematical Institute of Russian
Academy of Sciences,}\\
\textit{\small Gubkin st., 8, 119991, Moscow, Russia,}\\
\texttt{\small arefeva@mi.ras.ru, rgorbachev@mi.ras.ru}
\\
\textit{\small ${}^2$ Institute of Theoretical and Experimental
Physics,}\\
\textit{\small B.Cheremushkinskaya st. 25, Moscow, 117218,}\\
\texttt{\small pmedvedev@itep.ru}
\\
\textit{\small ${}^3$ Physics Department, Moscow State University,}\\
\textit{\small Moscow, Russia, 119899,}\\
\texttt{\small rogdin@gmail.com}}
\date{~}

\maketitle \thispagestyle{empty}

\vspace{1cm}
\begin{abstract}
We reexamine the oscillator level truncation method in the bosonic
String Field Theory (SFT) by calculation the descent relation
$\langle V_3|V_1\rangle=\tilde{Z}_3\langle V_2|$. For the ghost
sector we use the fermionic vertices in the standard oscillator
basis. We propose two new schemes for calculations. In the first one
we assume that the insertion satisfies the overlap equation for the
vertices and in the second one we use the direct calculations. In
both
 schemes we get the correct structures of the exponent and
 pre-exponent of the vertex $\langle V_2|$, but
 we find out different normalization factors $\tilde{Z}_3$.
\end{abstract}


\newpage
\section{Introduction}
The Witten cubic bosonic string field theory (CSFT) \cite{W}
exhibits very nice algebraic properties. Fields in this theory
belong to the infinite-dimension non-commutative
$\mathbb{Z}_2$-graded associative algebra. The action has the
Chern-Simons form

\begin{equation}
S=\frac 1 2\int \Phi\star Q\Phi+\frac g 3 \int \Phi\star \Phi\star
\Phi.
\end{equation}
Representations of Witten's algebra in the oscillator and conformal
languages were proposed \cite{GJ, LPP} (for a review see
\cite{Ohmori, ABGKM, Taylor:2003gn}).

To formulate Witten's theory in the oscillator representation it is
enough to construct only two vertices: $\langle |V_3 |\rangle$ which
represents $\star$, and $\langle V_1|$ which represents $\int$
\cite{GJ}. In this representation string fields $\Phi$ are realised
as ket-vectors $|\Phi\rangle$ in the Fock space. The vertices $V_1
,V_3$ are defined up to two normalization factors that can be
absorbed in to the charge and string field redefinitions.

The main difficulty in actual calculations with vertices $V_i$ is
that they are defined by infinite-dimensional matrices \cite{GJ,
CST,Samuel2,Ohta}. To perform real calculation one can use finite
dimensional approximation of these matrices \cite{T}. One calls this
method as the oscillator level truncation method.

An alternative method is the field level truncation method which
originates from \cite{KS,MT}. This method was used with great
success in refs. \cite{SZw,RSZ,ABKM2} for tachyon potential
calculations. The quartic term in the tachyon potential can be also
obtained from the Born zero momentum 4-point CSFT-diagram
\cite{T,KS}. It can be also computed within the oscillator level
truncation method and the result is in a good agreement with the
tachyon potential obtained within field level truncation method.

To justify more the oscillator level truncation method it is worth
to check within this method some exact relations. In \cite{KF} it
was proposed to check descent relations between string vertices.
Descent relations \cite{GJ} relate N-string vertices, $N\geq 1$ in a
natural way:
\begin{equation}
\label{descrel31} \langle V_{N+1}|V_1\rangle=\langle V_N|.
\end{equation}
In ref. \cite{KF,B} it was found that these relations has the form
$\langle V_3|V_1\rangle=\tilde{Z}_3\langle V_2|$, where
$\tilde{Z}_3\neq1$. The appearance $\tilde{Z}_N\neq1$ in descent
relations is considered as an anomaly. We'll discuss this
interpretation in section 2. In order to understand the origin of
the $\tilde{Z}_N\neq1$ breaking one has to investigate the vertex
descent relations in more details. It is true that one cannot a
priori expect $\tilde{Z}_N=1$ for arbitrary vertices $V_1$, $V_N$
and $V_{N+1}$. We present a self-consistence scheme to define
vertices $\langle \hV_N|, ~N\geq 2$ in terms of $\langle V_1|$ and
$\langle |V_3|\rangle$ which obey (\ref{descrel31}). For vertices
defined via overlap equations one can not expect (\ref{descrel31})
since solutions of overlaps are defined up to a number. Generally,
instead of (\ref{descrel31}) one has the descent relation of the
form
\begin{equation}
\langle V_{N+1}|V_1\rangle=Z_N Z_{-1}^{-1}Z_{N+1}^{-1}\langle V_N|.
\end{equation}
 This factorization was found in
\cite{Belov2} as a result of direct calculations with vertices in
continuous $\kappa$-basis \cite{Rastelli:2001hh}. Note that it is
more convenient to care out some calculations analytically when the
matrices are presented in the diagonal form in the continuous
$\kappa$-basis \cite{B}.

We verify the descent relation for $\langle V_2|V_1\rangle$ and
$\langle V_3|V_1\rangle$ in Section 3. In the first case the
calculation is similar to that of \cite{GJ} and gives
$\tilde{Z}_2=1$ without any truncation. In the second case we use
two schemes in the ghost sector, one (labeled by (1)) with an
overlap equation for the mid-point insertion and the second (labeled
by (2)) without this trick. Both our schemes differ from \cite{KF}
in the way we treat the ghost zero modes. These schemes give the
descent relation in the form
\begin{equation}
\langle V_3^m|V_1^m\rangle=\tilde{Z}^m_3(n)\langle
V_2^m|,\quad\langle
V_3^{gh}|V_1^{gh}\rangle^{(i)}=\tilde{Z}^{gh}_{3{(i)}}(n)\langle
V_2^{gh}|,\quad i=1,2
\end{equation}
where  $\tilde{Z}^m_3(n)$ and $\tilde{Z}^{gh}_3(n)$ are level $n$
dependent constants. For $n\to\infty$ the matter part
$\tilde{Z}^{m}_3(n)\to0$, but the ghost part
$\tilde{Z}^{gh}_3(n)\to\infty$ . Combining the matter sector with
the ghost sector we get the normalization $\tilde{Z}_3$ as
$\tilde{Z}_3(n)=\tilde{Z}^{gh}_3(n)\tilde{Z}^{m}_3(n)$. We
numerically prove that $\tilde{Z}_3(n)$ has a finite limit as $n\to
\infty$ in both our schemes. The surprising fact is that in two
different schemes of calculations we get two different limits,
namely $\tilde{Z}_{3(1)}(\infty)\cong0.070075$ in the scheme with
overlap and $\tilde{Z}_{3(2)}(\infty)\cong0.199381$ in the scheme
without overlap. The second result $\tilde{Z}_{3(2)}(\infty)$
coincides with the result obtained in \cite{KF}.

\section{$Set~Up$}
In the operator representation string fields $\Phi$ are realized as
ket vectors $|\Phi\rangle$ in the Fock space $\mathcal{H}$ of matter
and ghost modes. For the discussion below it is convenient to use a
graphic representation similar to a standard diagram technique used
to calculate vacuum expectation values of products of Wick
monomials. In this technique a Wick monomial is represented by an
arrow, so that creation operators correspond to the arrows going to
the left and annihilation operators --- to the arrows going to the
right. Here for simplicity we use one arrow for all creation or
annihilation operators of a string. So we represent bra vectors by a
point with the right arrows and ket vectors by a point with the left
arrows, i.e. we represent a state $|\Phi\rangle$ as

\hspace{6cm}
\begin{picture}(150,30)
\put(1,22){\text{$|\Phi\rangle$~=}} \put(24,23.5){\vector(-1,0){10}}
\put(23.85,22.45){\text{$\bullet$}}
\end{picture}

\vspace{-1.8cm}

To formulate Witten's theory \cite{W} it is necessary and sufficient
to introduce only two vertices:
$${}_{12}\langle |V_3 |\rangle_3 \in \mathcal{H}^*\otimes \mathcal{H}^*\otimes \mathcal{H}$$
which represents $\star$, and $$\langle V_1|\in \mathcal{H}^*$$
which represents $\int$.

In the graphic notations $\langle V_1|$ is represented by an arrow
outgoing from a circle to the right

\hspace{3cm}
\begin{picture}(150,30)
\put(35,22){\text{$\langle V_1|$~=}} \put(50.3,22.1){$\circ$}
\put(52,23.1){\vector(1,0){10}} 
\end{picture}
and $\langle V_1|\Phi\rangle$ is represented as

\hspace{3cm}
\begin{picture}(150,30)
\put(19.7,23.3){$\circ$} \put(21.5,24.3){\vector(1,0){10}}
\put(49.8,23.3){$\bullet$} \put(50,24.3){\vector(-1,0){10}}
\put(55,23.3){\text{~=~~}} \put(62,23.3){$\circ$}
\put(63.68,24.4){\line(1,0){10}}\put(73.5,23.3){$\bullet$}
\end{picture}
\vspace{-2.5cm}

The Witten vertex ${}_{12}\langle |V_3 |\rangle_3$ is represented by
three arrows outgoing from a common point. Two of them are going to
the right and one is going to the left. So we can display the star
product as follows:

\hspace{2.5cm}
\begin{picture}(150,40)
\put(0,22.5){\text{$|\Phi_2\rangle \star |\Phi_1\rangle ~=~$}}
\put(40,23.5){\vector(-1,0){10}} \put(40,23.5){\vector(1,1){7}}
\put(40,23.5){\vector(1,-1){7}}
\put(60,31){\vector(-1,0){7}} \put(60,30){$\bullet$}
\put(60,16.5){\vector(-1,0){7}} \put(60,15.5){$\bullet$}
\put(70,23){\text{$~=~$}}
\put(90,23.5){\vector(-1,0){10}} \put(90,23.5){\line(1,1){7}}
\put(90,23.5){\line(1,-1){7}}
\put(97,30){$\bullet$}\put(97,15){$\bullet$}
\put(21.5,5){\text{$=$}} \put(30,5){\text{${}_{12}\langle
|V_3|\rangle_3 |\Phi_2\rangle |\Phi_1\rangle$}}
\put(71.5,5){\text{$=$}}
\put(80,5){\text{$|\Phi_2\star\Phi_1\rangle_3 .$}}
\end{picture}

One can build an infinite tower of vertices ${}_{1..N}\langle
|V_{N+1}|\rangle_{N+1}$ by gluing of $N-1$ vertices $V_3$. This
 tower graphically has the form of a tree with $N+1$ arrows.
 Inside this tree we can glue free ends of the vertices $V_3$ arbitrary,
all of these gluing are equivalent (it follows from the
associativity of the star product \cite{GJ}) and we will use the
simplest case.

Using the vertex ${}_{1..N}\langle |V_{N+1}|\rangle_{N+1}$, one can
define the star product for $N$ fields as:
\begin{eqnarray}
\label{definition of N+1 vertex} {}_{1..N}\langle
|V_{N+1}|\rangle_{N+1}|\Phi_N\rangle ...|\Phi_1\rangle
\equiv|\Phi_N\rangle \star ...\star|\Phi_1\rangle.
\end{eqnarray}
The LHS of (\ref{definition of N+1 vertex}) can be displayed as

\hspace{1cm}
\begin{picture}(170,40)
\put(40,23.5){\vector(-1,0){10}} \put(40,23.5){\vector(1,1){7}}
\put(40,23.5){\vector(1,-1){7}}\put(40,23.5){\vector(3,1){7}}
\put(60,31){\vector(-1,0){7}} \put(60,30){$\bullet$}
\put(60,26){\vector(-1,0){7}} \put(60,25){$\bullet$}
\put(60,16.5){\vector(-1,0){7}} \put(60,15.5){$\bullet$}
\put(58,21){\text{$......$}}
\put(63,30){\text{$\Phi_1$}}\put(63,25){\text{$\Phi_2$}}\put(63,15.5){\text{$\Phi_N$}}
\put(70,23){\text{$~=~$}}
\put(90,23.5){\vector(-1,0){10}} \put(90,23.5){\line(1,1){7}}
\put(90,23.5){\line(3,1){7}} \put(90,23.5){\line(1,-1){7}}
\put(97,30){$\bullet$}\put(97,15){$\bullet$}\put(97,25){$\bullet$}
\put(100,30){\text{$\Phi_1$}}\put(100,25){\text{$\Phi_2$}}\put(100,15.5){\text{$\Phi_N$}}
\put(96,21){\text{$.......$}}
\end{picture}
\vspace{-1.5cm}

 One can also build an infinite tower of
vertices ${}_{1...N}\langle V_{N} |$ associated with constructed
above vertices ${}_{1..N}\langle |V_{N+1}|\rangle_{N+1}$ by adding
one more $\langle V_1|$ as shown below,

\hspace{4.5cm}
\begin{picture}(170,25)
\put(0,13.5){\vector(1,0){10}} \put(-1.5,12.5){\text{$\circ$}}
\put(25,13.5){\vector(-1,0){10}} \put(25,13.5){\vector(1,1){7}}
\put(25,13.5){\vector(3,1){7}} \put(25,13.5){\vector(1,-1){7}}
\put(29,11.5){\text{...}}
\put(37,12.5){\text{=}}
\put(45.2,12.5){\text{$\circ$}}\put(57,13.5){\line(-1,0){10}}
\put(57,13.5){\vector(1,1){7}} \put(57,13.5){\vector(3,1){7}}
\put(57,13.5){\vector(1,-1){7}} \put(61,11.5){\text{...}}
\end{picture}

\vspace{-0.5cm}

Let us call the set of these vertices as Witten's tower of vertices
and denote them by a hat. For consistency we will also put the hat
on the initial vertices $\langle V_1|$ and $\langle |V_3|\rangle$.

Witten's tower is defined by $\langle \hat{V}_1 |$ and $\langle
|\hat{V}_3|\rangle$ which are defined up to a number.
 Therefore one needs two
constraints to fix Witten's tower
 unambiguously. The most natural way is to postulate  SFT-CFT \cite{LPP}
correspondence for two vertices:
\begin{equation}
\label{SFT-CFT corr.}\langle \hat{V}_N|\Phi_1\rangle
...|\Phi_N\rangle\equiv\langle
\Phi_1...\Phi_N\rangle_{\Sigma_N},\quad N=2,3.
\end{equation}
where $\Sigma_N$ is an appropriate Riemann surface \cite{LPP}. It is
unclear does this fixing of ambiguities also guarantee that
(\ref{SFT-CFT corr.}) holds for all vertices ($N>3$) in the tower.

We find also ``ket'' $|\hat{V}_1\rangle$ as a solution of the
``descent relation'' \cite{GJ}
\begin{equation}
\label{V_1} \langle\hat V_2|\hat V_1\rangle=\langle \hat V_1|.
\end{equation}

This defining equation for $|\hat{V}_1\rangle$ can be represented
graphically. We display a vertex corresponding to $|\hat V_1\rangle$
by an arrow outgoing from $\ast$ to the left. Wick theorem gives
that the LHS of (\ref{V_1}) can be presented as

\hspace{5cm}
\begin{picture}(170,20)
\unitlength=0.8mm
\put(10,15){\vector(1,1){7}} \put(10,15){\vector(1,-1){7}}
\put(0,13.75){\text{$\circ$}}\put(10,15){\line(-1,0){8}}
\put(30,8){\vector(-1,0){10}}
\put(28.5,6.7){\text{$*$}}\put(38,14){\text{$=$}}
\put(47.8,13.75){\text{$\circ$}}\put(60,15){\line(-1,0){10}}
\put(60,15){\vector(1,1){7}} \put(60,15){\line(1,-1){7}}
\put(65.6,6.8){\text{$*$}}
\end{picture}
\vspace{-0.5cm}
\\
In this notations equation (\ref{V_1}) looks like

\hspace{5.5cm}
\begin{picture}(170,25)
\unitlength=0.8mm
\put(10,15){\line(-1,0){8}} \put(10,15){\vector(1,1){7}}
\put(10,15){\line(1,-1){7}}
\put(0,13.5){\text{$\circ$}}\put(16,7){\text{$*$}}
\put(28,14){\text{$=$}}
\put(40,15){\vector(1,0){10}} \put(38,13.5){\text{$\circ $}}
\end{picture}
\vspace{-0.5cm}

With $|\hat{V}_1\rangle$ subject to (\ref{V_1}) we are going to
verify the descent relations
\begin{equation}
\label{DescRel} \langle\hat V_{N+1}|\hat V_1\rangle=\langle\hat
V_{N}|.
\end{equation}
It can be represented by the graphs for $N>1$ as

\hspace{2.5cm}
\begin{picture}(170,25)
\unitlength=0.8mm
\put(41,22){\text{$1$}} \put(41,2){\text{$N+1$}}
\put(30,13.5){\vector(1,1){10}} \put(30,13.5){\vector(3,1){10}}
\put(30,13.5){\vector(1,-1){10}} \put(38,12){\text{...}}
\put(30,13.5){\vector(2,-1){10}}\put(17.8,12.35){\text{$\circ$}}
\put(30,13.6){\line(-1,0){10}}
\put(58,9){\vector(-1,0){10}} \put(56.6,7.7){\text{$*$}}
\put(68,12){\text{$=$}}
\put(90,13.5){\vector(1,1){10}} \put(90,13.5){\vector(3,1){10}}
\put(90,13.5){\vector(1,-1){10}}
\put(98,12){\text{...}}\put(77.8,12.2){\text{$\circ$}}
\put(90,13.6){\line(-1,0){10}} \put(101,2){\text{$N$}}
\put(101,22){\text{$1$}}
\end{picture}
\\
To prove (\ref{DescRel}) we have to use our construction for
vertices $\langle \hat{V}_{N+1}|$ (as we have mentioned above, we
can take a simplest graph), namely

\hspace{4.5cm}
\begin{picture}(170,30)
\put(0,13.5){\text{$\langle \hat{V}_{N+1}|$}}
\put(20,13.5){\text{$=$}}
\put(30,14.5){\vector(1,0){40}} \put(28.5,13.5){\text{$\circ$}}
\put(40,14.5){\vector(1,1){10}} \put(55,18){\text{...}}
\put(45,14.5){\vector(1,1){10}}\put(60,14.5){\vector(1,1){10}}
\put(34,20.5){\text{$N+1$}}\put(54,20.5){\text{$N$}}
\put(70,20.5){\text{$2$}} \put(67,10.5){\text{$1$}}
\end{picture}
\vspace{-1cm}
\\
Now

\hspace{0.5cm}
\begin{picture}(170,25)
\put(0,13.5){\text{$\langle \hat{V}_{N+1}|\hat{V}_1\rangle$}}
\put(20,13.5){\text{$=$}}
\put(27,14.5){\vector(1,0){40}} \put(25.4,13.5){\text{$\circ$}}
\put(37,14.5){\vector(1,1){10}} \put(52,18){\text{...}}
\put(42,14.5){\vector(1,1){10}}\put(57,14.5){\vector(1,1){10}}
\put(82,14.5){\vector(-1,0){10}}\put(81,13.5){\text{$*$}}\put(87,13.5){\text{$=$}}
\put(97,14.5){\vector(1,0){40}}
\put(95.5,13.5){\text{$\circ$}}\put(115.9,23.5){\text{$*$}}
\put(107,14.5){\line(1,1){10}} \put(122,18){\text{...}}
\put(112,14.5){\vector(1,1){10}}\put(127,14.5){\vector(1,1){10}}
\end{picture}
\vspace{-1cm}
\\

Taking into account (\ref{V_1}) we get that

\hspace{1.5cm}
\begin{picture}(170,30)
\put(10,14.5){\vector(1,0){40}} \put(8.5,13.5){\text{$\circ$}}
\put(20,14.5){\line(1,1){10}} \put(35,18){\text{...}}
\put(25,14.5){\vector(1,1){10}}\put(40,14.5){\vector(1,1){10}}
\put(29.5,24){\text{$*$}} \put(47,10.5){\text{$1$}}
\put(50,20.5){\text{$2$}}\put(34,20.5){\text{$N$}}
\put(16,20.5){\text{$N+1$}}
\put(62.5,13.5){\text{$=$}}
\put(80,14.5){\vector(1,0){40}} \put(78.5,13.5){\text{$\circ$}}
\put(105,18){\text{...}}
\put(95,14.5){\vector(1,1){10}}\put(110,14.5){\vector(1,1){10}}
\put(117,10.5){\text{$1$}} \put(119,19.5){\text{$2$}}
\put(97,19.5){\text{$N$}}
\end{picture}
\vspace{-1.5cm}
\\
that exactly gives $\langle \hat{V}_N|$.

 Bra vertices $\langle V_i|$  are usually found as the solutions of
 the overlap equations \cite{GJ,CST,Ohta,Samuel:1987fi}. However,
these solutions are defined up to a numerical factors, i.e. the
vertex $\langle\hat V_N|$ in Witten's tower differs from any other
vertex $\langle V_N|$ by a number $\langle\hat V_N|=Z_N\langle V_N|$
and $|\hat V_1 \rangle=Z_{-1} | V_1 \rangle$. Hence the descent
relations for $\langle V_N|$ look like
\begin{equation}
\label{DR_with_Z} \langle
V_{N+1}|V_1\rangle=Z_NZ_{-1}^{-1}Z_{N+1}^{-1}\langle V_N|.
\end{equation}

The factorization of coefficient in descent relation was first found
in \cite{Belov2}, where vertices were built in the continuous
$\kappa$ - basis. It turned out that descent relation contains the
coefficient not equal to one, and moreover this coefficient admits a
factorization on three factors. Each of these factors was
interpreted as boundary entropy of CFT. Using this factorization in
\cite{Belov2} the vertices were redefined as $\langle\langle
V_N|=Z_N\langle V_N|$.
 After this redefinition the descent relation has the coefficient equal to
 one, $\langle\langle V_{N+1}|V_1\rangle\rangle=\langle\langle V_N|$.

\section{Descent Relations for $N=1$ and $N=2$}
\subsection{$\langle V_2|V_1\rangle$}
The vertices which will be used were built in \cite{GJ, CST,
Samuel2, Samuel:1987fi, Taylor:2003gn}. Here and below we use the
vectors notations for the modes of the matter fields
$X^{\mu}(\sigma)$ as\footnote{We suppress the space-time index
below.} $a=(a_0,a_1,...), a^{*}=(a_0^*, a_{-1},...)$ and for the
ghost fields $b(\sigma),~c(\sigma)$ as $b, b^{*}$ and $c, c^{*}$
similarly (see Appendix). In some cases we will include the zero
mode $b_0$ into $b$ with a special comment. The zero mode $c_0$
enter neither $c$ nor $c^*$. Also we use the matrix
$S_{nk}=(-)^{n+1}\delta_{nk}$ and here $n,k\geq 0$.

We start with the matter sector. The explicit oscillator
representation of $V_1$ and $V_2$ are given by
\begin{equation*}
 |V_1^{m}\rangle=\exp\left\{\frac12a^{*1}
Sa^{*1}\right\}|0\rangle_1,\, \langle
V_2^{m}|={}_{1,2}\langle0|\exp\left\{ a^1Sa^2\right\}, \, \langle
V^{m}_1|={}_1\langle 0|\exp\left\{\frac 1 2 a^1Sa^1 \right\}.
\end{equation*}

The LHS of (\ref{DR_with_Z}) is
\begin{equation*}
\langle V_2^{m}|V_1^{m}\rangle= {}_{1,2}\langle0|
e^{a^1Sa^2}e^{\frac12a^{1*}Sa^{1*}}|0\rangle_1=
{}_2\langle0|e^{\frac12a^2SSSa^2}=
{}_{2}\langle0|\exp\left\{\frac12a^2Sa^2\right\}=\langle V_1^{m}|.
\end{equation*}
Here the Wick theorem \cite{KP} is used.

Now we consider the descent relation in the ghost sector. For this
calculations we take the following vertices (in this case
$n,m\geq1$):
\begin{eqnarray*}
|V_1^{gh}\rangle=\frac
i4b_+^1b_-^1\exp\left\{b^{*1}Sc^{*1}\right\}|+\rangle_1,\quad
\langle
V_1^{gh}|={}_1\langle+|\exp\{-b^1Sc^1\}b_-^1b_+^1 \frac i4,\nonumber\\
\langle V_2^{gh}|={}_{1,2}\langle+|(b_0^1-b_0^2)
\exp\left\{-b^1Sc^2-b^2Sc^1\right\}.
\end{eqnarray*}
The mid-point insertions are
$$b_+\equiv
b(\pi/2)=\sum_{-\infty}^\infty i^nb_n,\qquad b_-\equiv
b(-\pi/2)=\sum_{-\infty}^\infty i^{-n}b_n.$$

Let us write down the LHS of the descent relation
\begin{equation*}
\langle
V_2^{gh}|V_1^{gh}\rangle={}_{1,2}\langle+|(b_0^1-b_0^2)e^{-b^1Sc^2-b^2Sc^1}
\frac i4b_+^1b_-^1e^{b^{1*}Sc^{1*}}|+\rangle_1.
\end{equation*}
We use the overlap equation for $b_+$ and $b_-$ to change the index
of insertions from ``1'' to ``2''. To exchange the order of the
vacuum states we have to take into account their parity, i.e.
$${}_{1,2}\langle +|=-{}_{2,1}\langle +|.$$
So using \cite{KP} we have
\begin{equation}
\langle
V_2^{gh}|V_1^{gh}\rangle=-{}_{2}\langle+|e^{-b^2SSSc^2}b_+^2b_-^2\frac
i4=-{}_2\langle+|\exp\{-b^2Sc^2\}b_+^2b_-^2\frac i4=\langle
V_1^{gh}|.
\end{equation}

Hence for the vertices which include the matter and ghost sectores
the descent relation with coefficient $\tilde{Z}_2$ equals to one
has the form
\begin{equation}
\langle V_2|V_1\rangle=\langle V_1|.
\end{equation}
We can conclude that the vertices $\langle V_1|$, $\langle V_2|$ and
$|V_1\rangle$
 can be considered as
the vertices from some Witten's tower.

\subsection{$\langle V_3|V_1\rangle$}

In this subsection we will consider the descent relation for
vertices $\langle V_3|$ and $|V_1\rangle$. This calculation is not
so simple as above.

We perform our calculation of the descent relation starting with the
ghost sector (notations are listed in Appendix)
\begin{equation}
\langle V_3^{gh}|V_1^{gh}\rangle=  N_3\frac i 4{}_{321}\langle
+|e^{- b^r X^{rs}c^s}b_+^1b_-^1e^{b^{*1}Sc^{*1}}|+\rangle_1.
\label{spusk1}
\end{equation}
Here we include the zero mode $b_0$ into $b$. In calculations we
drop the coefficient $\frac i 4 N_3=\frac i 4\frac{3\sqrt{3}}{4}$,
it will be restored in the final expression.

We will explore two schemes to evaluate (\ref{spusk1}). We can take
the insertion $b_+^1b_-^1$ as it stands or we can use the overlap
equation for $\langle V_3^{gh}|$
\begin{equation}
\langle V_3^{gh}|(b^{i-1}(\sigma)-b^i(\pi-\sigma))=0,\qquad i=1,2,3
\end{equation}
to move $b_+^1$ and $b_-^1$ to the Fock spaces with index ``2'' or
``3''. The first scheme of calculations will be presented in Section
3.2.2 and the second one in the Section 3.2.1.

In \cite{KF} there was presented another scheme where the vertex
$|V_1\rangle$ was considered in $|-\rangle$ vacuum and the direct
calculations were performed.

\subsubsection{``Overlap Calculations'' in Ghost Sector}
We use the overlap equation to change the index of insertion $b_+$
from ``1'' to ``2'' and $b_-$ from ``1'' to ``3''. So the expression
we have to evaluate reads (here $b^r$ include $b_0^r$):
\begin{equation}
\label{V3V1-ish}{}_{321}\langle
+|e^{-b^rX^{rs}c^s}e^{b^{*1}Sc^{*1}}|+\rangle_1 b_+^2b_-^3.
\end{equation}
Separating the zero mode $b_0^1$ we get
\begin{eqnarray*}
{}_{32}\langle+|e^{-W_2}{}_{1}\langle
+|(1-b_0w)e^{-W_1}e^{b^{*1}Sc^{*1}}|+\rangle_1
b_+^2b_-^3=-{}_{32}\langle+|e^{-W_2}{}_{1}\langle
-|e^{-W_1}we^{b^{*1}Sc^{*1}}|+\rangle_1b_+^2b_-^3.
\end{eqnarray*}
Here we put:
$$
w=X^{11}_{0m}c_m^1+X^{1p}_{0m}c^p_m,\quad p=2,3
$$
in $W_1$ we have collected the dependence on non-zero modes
$b^1,\,c^1$ and $W_2$ depends on the modes of the second and on the
third strings (see Appendix).

The zero mode dependence
 will be of a special interest. In our calculations for any matrix
  $A_{nm}$ we denote $\bar A_k\equiv A_{0k}$
and $\hat A_{nm}\equiv A_{nm}$ with $n>$0.

We need to calculate the following expression
\begin{equation}
\label{average} {}_{1}\langle
-|e^{-W_1}we^{b^{*1}Sc^{*1}}|+\rangle_1.
\end{equation}
To evaluate \eqref{average} we use the generalized formula of eq.
(19) in \cite{KP} given in \cite{ Kishimoto:2001ac,KF}. It has a
form:
\begin{eqnarray}
\label{sq-states} \langle -| \exp(-bXc+b\lambda^c+\lambda^b
c)\exp(b^*Sc^*+b^*\mu^c+\mu^b c^*) |+\rangle=\nonumber\\
=\det(1-SX)\exp\left\{\mu^b\frac{1}{(1-XS)}(X\mu^c-\lambda^c)+\lambda^b
\frac{1}{1-SX}(\mu^c-S\lambda^c)\right\}.
\end{eqnarray}
Here $\lambda^b, \lambda^c$ and $\mu^b, \mu^c$ are anticommuting
vectors.

So using eq. (\ref{sq-states}) and
${}_{3,2}\langle+|=-{}_{2,3}\langle+|$ we can evaluate
(\ref{average}). Thereby the expression (\ref{V3V1-ish}) gives
\begin{eqnarray}\label{spusc1otvet}
\langle V_3^{gh}|V_1^{gh}\rangle=-\det(Z^{-1}){}_{23}\langle
+|\bU^{1q}c^qe^{b^pU^{pq}c^q}b_+^2b_-^3,
\end{eqnarray}
where
\begin{equation}
\label{U}
U^{pq}_{nm}=-X^{p1}_{nk}(ZS)_{kl}X^{1q}_{lm}-X^{pq}_{nm},\quad
\text{here}\quad n\geq0,\quad m,k,l\geq1,\quad p,q=2,3.
\end{equation}
Here we defined
$$ Z=(1-SX)^{-1},$$
where $X_{nm}\equiv X^{11}_{nm}$, and we stress that $n,\,m\geq1$.

To put the expression (\ref{spusc1otvet}) to a normal form we move
the insertion $b_+^2b_-^3$  to the left to the vacua through the
exponent and pre-exponential factor. We get the following
pre-exponential factor (for some notations see Appendix):
\begin{eqnarray}
\label{*} (b^2K+b_0^2+b_0^p\bU^{p2}L+b^p\hU^{p2}L)(b^3L+b_0^3+
b_0^p\bU^{p3}K+b^p\hU^{p3}K)\bU^{1q}c^q-\nonumber\\
-(b^2K+b_0^2+b_0^p\bU^{p2}L+b^p\hU^{p2}L)\bU^{13}K+
(b^3L+b_0^3+b_0^p\bU^{p3}K+b^p\hU^{p3}K)\bU^{12}L.
\end{eqnarray}
Because the vacua stays on the left the items with creation
operators are dropped.

The matrices $\hat U^{pq}$ and rows $\bar U^{pq}$
 for different $p$-s and $q$-s are not independent. By the definition
 (\ref{U}) and properties
 of the Neumann matrix
 (see Appendix) we have:
\begin{eqnarray}
\label{U-hat-properties}
\hU^{i2}+\hU^{i3}=-S,&\,\,&\hU^{2i}+\hU^{3i}=-S,\qquad i=2,3,
\end{eqnarray}
\begin{eqnarray}
\label{U-bar-properties}
\bU^{1j}+\bU^{2j}+\bU^{3j}=0,&&\bU^{j2}+\bU^{j3}=0,\qquad j=1,2,3.
\end{eqnarray}
For example one can prove the first property \cite{KF}
\begin{eqnarray*}
\hat U^{i2}+\hat U^{i3}&=&-X^{i2}-X^{i1}(1-SX^{11})^{-1}SX^{12}-
X^{i3}-X^{i1}(1-SX^{11})^{-1}SX^{13}\\
&=&-X^{i2}-X^{i3}-X^{i1}(1-SX^{11})^{-1}S(X^{12}+X^{13})\\
&=&-X^{i2}-X^{i3}-X^{i1}(1-SX^{11})^{-1}S(S-X^{11})\\
&=&-X^{i2}-X^{i3}-X^{i1}(1-SX^{11})^{-1}(1-SX^{11})=-S.
\end{eqnarray*}
By using (\ref{U-hat-properties}) an expression (\ref{*}) can be
simplified. We express all the matrices in terms of the matrix
$\hU^{22}$.

It is useful to introduce the following notations
\begin{eqnarray}
\alpha_2=(b^2+b^3)K+(b^2-b^3)\hU^{22}L,&\quad&
\beta_2=b_0^2(1+\bar{U}^{22}L)+b_0^3\bar{U}^{32}L,\nonumber\\
\alpha_3=(b^2+b^3)L-(b^2-b^3)\hU^{22}K,&\quad&
\beta_3=b_0^3(1+\bar{U}^{33}K)+b_0^2\bar{U}^{23}K.
\end{eqnarray}
Thereby, the result of moving the $b_+b_-$ insertion to the left
(pre-exponential factor (\ref{*})) can be presented in the following
form
\begin{eqnarray}
\label{predexp} (\alpha_2+\beta_2)(\alpha_3+\beta_3)\bU^{1q}c^q-
(\alpha_2+\beta_2)\bU^{13}K+(\alpha_3+\beta_3)\bU^{12}L.
\end{eqnarray}
So we have the following result
\begin{equation}
-\det(Z^{-1}){}_{23}\langle+|[(\alpha_2+\beta_2)(\alpha_3+\beta_3)\bU^{1q}c^q-
(\alpha_2+\beta_2)\bU^{13}K+(\alpha_3+\beta_3)\bU^{12}L]e^{b^pU^{pq}c^q}
\end{equation}
Prima facie it does not match the vertex $\langle V_2^{gh}|$,
because this expression does not have the correct structure of vacua
and exponent, but it has big pre-exponent and has not correct
structure in exponent. Further we'll show the way to get the desired
expression.

Below the following relations which are verified numerically (they
hold true form the first level) will be useful
\begin{equation}\label{svoistvaU}
\bU^{23}=-\bU^{32}S,\quad \bU^{22}=-\bU^{33}S,\quad
\bU^{12}=-\bU^{13}S.
\end{equation}

As a result we have:
\begin{equation}
\bU^{32}L=\bU^{23}K,\quad\bU^{22}L=\bU^{33}K,\quad\bU^{13}K=\bU^{12}L,\quad\bU^{32}L=-\bU^{22}K.
\label{chslenno}
\end{equation}
In (\ref{predexp}) we can find the column $\hU^{22}(L+K)$. For
further calculations we need to investigate this object numerically.

Substituting the explicit form of L and K we get the elements of the
column as:

$$a_m=2\sum_{k=0}^\infty(-)^k\hU^{22}_{m,2k}.$$

For any level $n$ one can calculate partial sums
$a_m(n)=2\sum_{k=0}^n(-)^k\hU^{22}_{m,2k}$ with $m\leq n$. For any
fixed $m$ we get a sequence of partial sums $a_m(n)$ as we vary n.
We want to prove that $a_m(n)\to 0$ as $n\to \infty$. We prove it
for $m=2,10,20,30,...,100$ taking the levels  $100 \leq n\leq 300$
with a step equal to 10. The results are plotted at the Figure. We
also find fits for the partial sums as
$|a_m(n)|\sim\frac{p_m}{n^{q_m}}$ (see Table). Due to the behavior
of the fits we can conclude that $|a_m(n)|\to0$ for $n\to\infty$
and, as a consequence, $a_m\to0$.

\begin{figure}[h!]
\begin{center}
\includegraphics[scale=0.9]{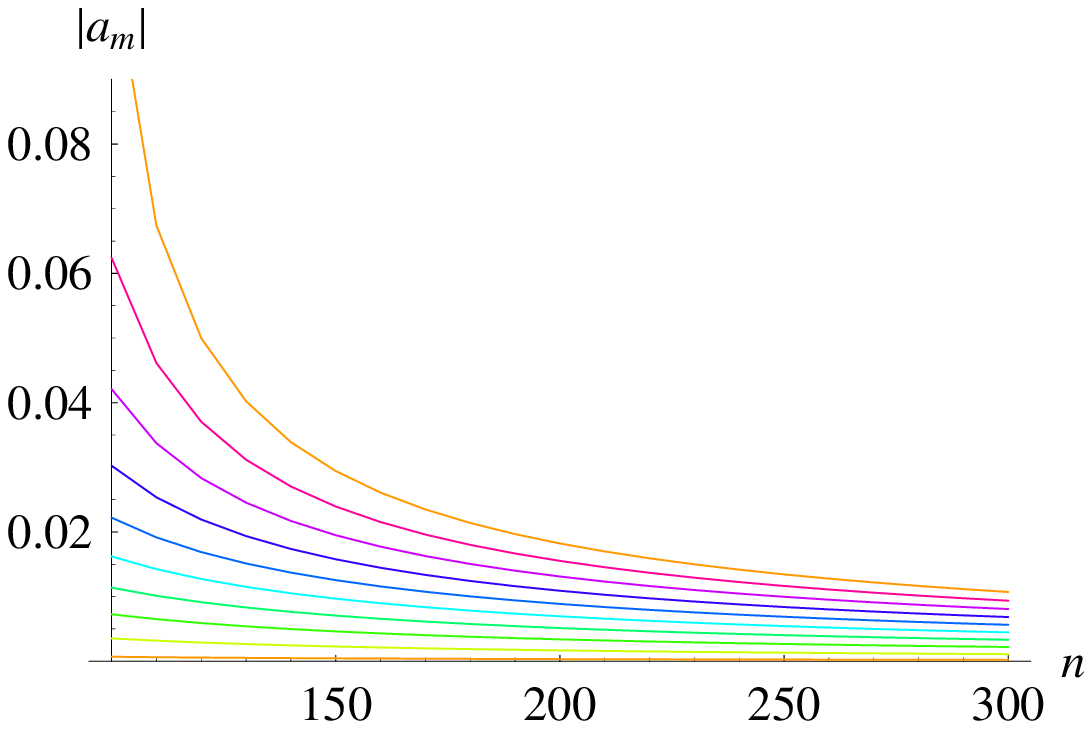}
\caption{} \label{U22}
\end{center}
\end{figure}
\begin{small}
\begin{table}[h]
\begin{center}
\caption{} \small{
\begin{tabular}{|c|c|c|c|c|c|c|c|c|c|c|c|}
  \hline
   m & 2 & 10 & 20 & 30 & 40 & 50 & 60 & 70 & 80 & 90 & 100 \\
  \hline
  p & 0.089 & 0.4617 & 1.0258 & 1.844 & 3.2117 & 5.806 & 11.463 & 26.227 & 76.781 & 364.31 & 16461 \\
\hline
  q & 1.053 & 1.059 & 1.077 & 1.109 & 1.156 & 1.221 & 1.309 & 1.428 & 1.595 & 1.854 & 2.529 \\
  \hline
\end{tabular}}
\end{center}
\end{table}
\end{small}
Due to the information from the Table 1 and Figure \ref{U22} we can
see that the speed of falling of the elements $a_m$ are increased
with extending of $m$. So we have that
\begin{equation}\label{chislenno2}
\hU^{22}(L+K)=0.
\end{equation}

Let us return to (\ref{predexp}). Substituting $\alpha$ and $\beta$
into (\ref{predexp})
 we get the expression for the terms linear in $b_0$.
Taking (\ref{chslenno}) and (\ref{chislenno2}) into account we get
\begin{eqnarray}
\label{linear-term} (b_0^2\left[J_0+(b^2\bar A_2+b^3\bar
A_3)\bU^{1q}c^q \right]-b_0^3\left[J_0+(b^2\bar B_2+b^3\bar
B_3)\bU^{1q}c^q \right]),
\end{eqnarray}
here (we express all $\bU^{12}$ and $\bU^{22}$ via eq.
(\ref{chslenno}))
\begin{eqnarray}
J_0&=&-(1+\bU^{22}(K+L))\bU^{12}L,\nonumber\\
\bar A^i&=&(-)^i\hU^{22}L(1+\bU^{22}(K+L))+L(1+\bU^{22}L)+K\bU^{22}K,\\
\bar
B^i&=&(-)^i\hU^{22}L(1+\bU^{22}(K+L))+K(1+\bU^{22}L)+L\bU^{22}K.\nonumber
\end{eqnarray}

So we see that $J_0$  contributes to $\tilde{Z}^{gh}_3(n)$. The
remaining terms which are quadratic in $b_0$ are presented in the
form:
\begin{eqnarray}
\label{quadr-term}
\beta_2\beta_3\bU^{1q}c^q=b_0^2b_0^3\left[(1+\bU^{22}L)^2-(\bU^{22}K)^2
\right]\bU^{1q}c^q.
\end{eqnarray}
The free part in (\ref{predexp}) contains of two terms:
$\alpha_2\alpha_3\bU^{1q}c^q$ and $(\alpha_3-\alpha_2)\bU^{12}L$. we
have to extract the normalization factor $J_0$ After this rescaling
these quantities vanish. On the figures we see the behavior of these
two normalized terms for $n\to\infty$.

\begin{figure}[h]
\begin{center}
\includegraphics[scale=0.8]{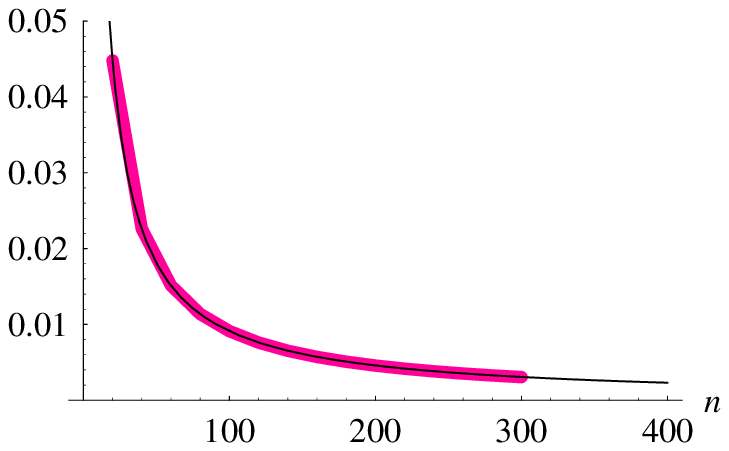}
\quad\includegraphics[scale=0.8]{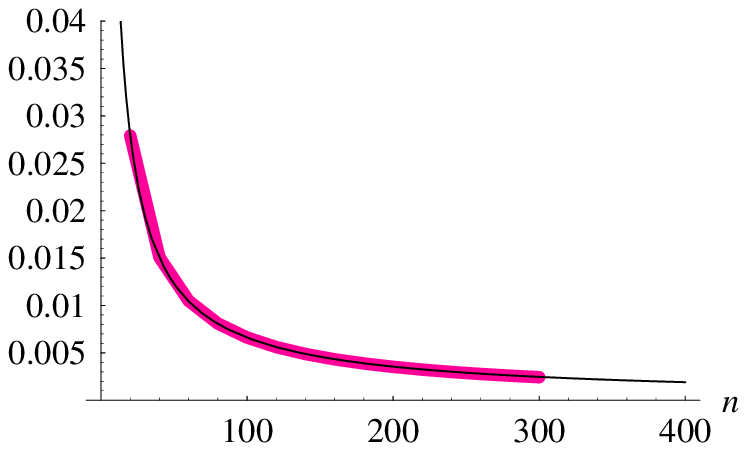} \caption{}
\label{free-terms}
\end{center}
\end{figure}
We find that for $n$=20...300 this curves are described by the fits
$\sim\frac{0.87}{n^{0.99}}$ and $\sim\frac{0.41}{n^{0.9}}$
respectively. So these terms vanish for $n\to\infty$.

Than we are going to rise up the terms linear and quadratic in $b_0$
term to the exponent. In this way we want to get the necessary
structure in the exponent.
\begin{equation}
{}_{23}\langle+|J_0(...)e^{bUc}\Longrightarrow
{}_{23}\langle+|J_0(b_0^2-b_0^3)e^{bVc}e^{bUc},
\end{equation}
(here the zero modes are contained in the matrix $V$ as well as in
the matrix  $U$). To get the desired structure the following
conditions has to take place (for $n\to \infty$)
\begin{equation}
\label{**} \bV^{pq}+\bU^{pq}\to0,
\end{equation}
\begin{equation}\label{conditions}
\hV^{pp}+\hU^{pp}\to0,\quad\hV^{p,p+1}+\hU^{p,p+1}\to -S.
\end{equation}
The constraint (\ref{**}) cancels the total zero modes dependence in
the exponent and the constraint
 (\ref{conditions}) will give the required structure of quadratic
 form. Below we'll discuss the realization of this program.

For convenience we'll go backwards, i.e. we will expand
$\exp\{bVc\}$ and compare the obtained structure with given by
(\ref{predexp}). In this way we will get conditions to rise up the
pre-exponential factor to the exponent. In our argumentations it is
important that we suppose the desired matrix $\hat{V}$ to be a
tensor product of two vectors linear in ghost modes, i.e.
$\hat{V}=\bV_1\otimes \bV_2$. Due to this guess the exponent
$e^{bVc}$ expands up to linear term (the quadratic term equals to
zero due to their Grassman structure) only.

Hence we get the following system of equations for unknown matrices
$V^{pq},~p,q=2,3$:
\begin{subequations}
\label{alpha}
\begin{eqnarray}
&1.&J_0(\bV^{2q}+\bV^{3q})=\left[(1+\bU^{22}L)^2-(\bU^{22}K)^2
\right]\bU^{1q}, \label{alpha-1}\\
&2.&(\bV^{3q}+\bV^{2q})c^q\cdot b^p\hV^{pq}c^q=0,\label{alpha-2}\\
&3.&\hV^{pq}=\frac{\bar{A}^p}{J_0}\otimes\bU^{1q},\label{alpha-3}\\
&4.&\hV^{pq}=\frac{\bar{B}^p}{J_0}\otimes\bU^{1q}.\label{alpha-4}
\end{eqnarray}
\end{subequations}
This equation do not define the matrices $V^{pq}$ completely  but
it's enough to rise up the pre-exponential factor to the exponent.

\begin{figure}[h!]
\begin{center}
\includegraphics[scale=0.8]{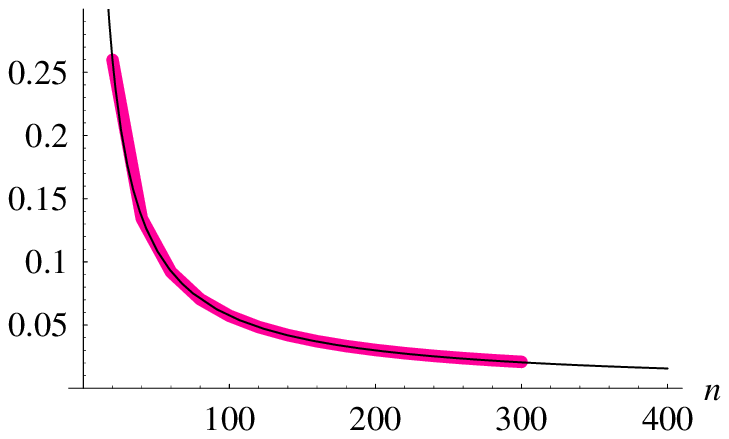}
\caption{} \label{consist}
\end{center}
\end{figure}

We can mark out two consistency conditions of the system:
\begin{equation}
\label{cons-cond}
\frac{\bar{A}^p}{J_0}-\frac{\bar{B}^p}{J_0}=T\frac{(1+\bU^{22}T)}{J_0}\to0,\quad
p=2,3.
\end{equation}
We calculated the LHS of (\ref{cons-cond})for $n=20...300$ and got
the fits for this set of points as $\frac{4.3}{n^{0.94}}$ (see
Figure \ref{consist}).

The equation \eqref{alpha-3} defines $\hV^{pq}$ and equation
\eqref{alpha-1} defines the sum $\bV^{2q}+\bV^{3q}$. The constraints
\eqref{alpha-2} follow from \eqref{alpha-1} and \eqref{alpha-3}:
$$
\frac1J_0\left[(1+\bU^{22}L)^2-(\bU^{22}K)^2 \right]\bU^{1q}c^q\cdot
b^2\frac{\bar{A}^p}{J_0}\otimes\bU^{1q}c^q=0.
$$

\begin{figure}[h!]
\begin{center}
\includegraphics[scale=0.8]{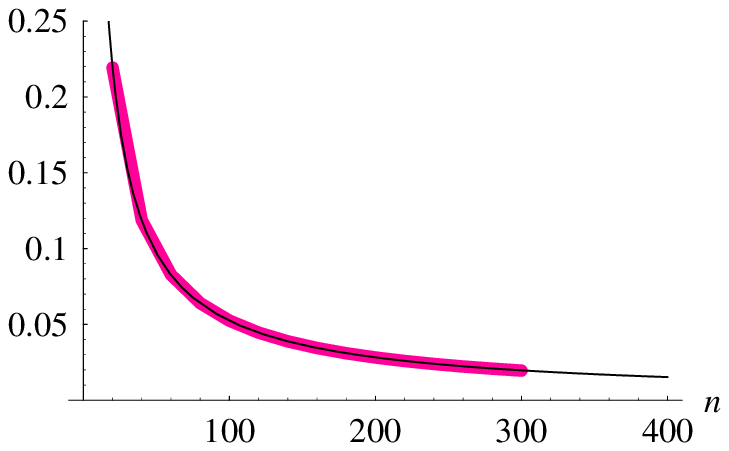}
\caption{} \label{UU22VV22}
\end{center}
\end{figure}

The equation \eqref{alpha-1} does not define $\bV^{pq}$-s separately
but only the sum. Let us put $\bV^{pq}\equiv-\bU^{pq}$ thus eq.
(\ref{**}) is fulfilled. Hence, we have to check  the constraint
\eqref{alpha-1} for this choice:
\begin{equation}
\label{***}
-J_0(\bU^{2q}+\bU^{3q})-[(1+\bU^{22}L)^2-(\bU^{22}K)^2]\bU^{1q}=0 .
\end{equation}
Due to (\ref{U-bar-properties}) we cancel the factor $\bU^{1q}$ in
both sides of eq. (\ref{***}) and have to check numerically the
equality
\begin{equation}
\label{****} J_0-[(1+\bU^{22}L)^2-(\bU^{22}K)^2]=0 .
\end{equation}
We checked it up to level 300 and got the fit as
$\frac{3.16}{n^{0.89}}$ (see Figure \ref{UU22VV22}).

\begin{figure}[h!]
\begin{center}
\includegraphics[scale=0.8]{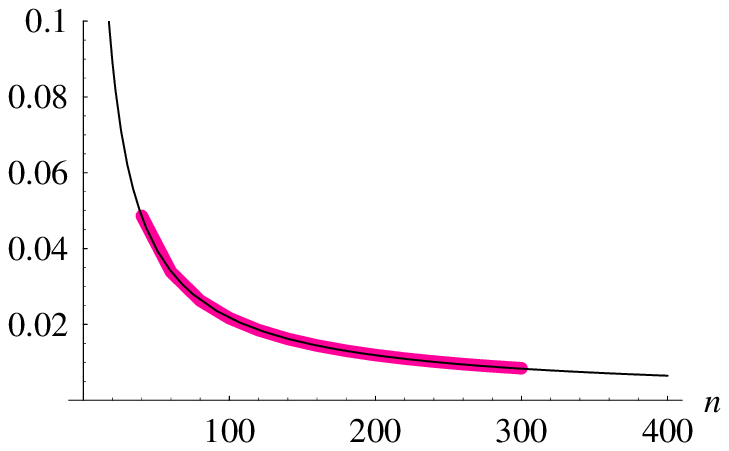}
\caption{} \label{U22+V22}
\end{center}
\end{figure}

The eq. \eqref{alpha-3} defines the matrices $\hV^{pq}=J_0^{-1}\bar
A^{p}\otimes\bU^{1q}$. So we have to check the conditions
(\ref{conditions}) for this choice. We do it up to level $n=300$.
For $p=2$ for the first condition in (\ref{conditions}) we have the
fit $\sim\frac{0.86}{n^{0.82}}$ (for the second condition in
(\ref{conditions}) and for $p=3$ the calculations and the fits are
almost the same). The result for numerical calculations and the fit
can be seen on the Figure \ref{U22+V22}.

Thereby, pre-exponential factor  (\ref{predexp}) rises up to the
exponent. These complete the proof of desired structure of the
exponential.

Hence we got that the descent relation in the ghost sector has the
following form
\begin{eqnarray}
{}_{321}\langle V_3^{gh}|V_1^{gh}\rangle_1&=&-J_0 N_3\frac i
4\det(1-SX){}_{23}\langle +|
(b_0^2-b_0^3)e^{-b^2Sc^3-b^3Sc^2}\nonumber\\&=& -J_0 N_3\frac i
4\det(1-SX){}_{23}\langle V_2| \equiv
\tilde{Z}_{ov}^{gh}{}_{23}\langle V_2^{gh}|.
\end{eqnarray}
Here we restore the coefficients which were dropped above.

\subsubsection{``Non-Overlap Calculations'' in Ghost Sector}

Here we examine the descent relation by straightforward calculations
without using overlaps.

The starting expression is eq. (\ref{spusk1})
\begin{equation}
\label{spusk2}
{}_{321}\langle+|e^{-b^rX^{rs}c^s}b_+^1b_-^1e^{b^*Sc^*}|+\rangle_1.
\end{equation}
It's useful to rewrite the insertion as
\begin{equation}
\label {bb} b_+^1b_-^1\equiv A+b_0^1B,
\end{equation}
here we denote (see Appendix)
\begin{eqnarray*}
A&\equiv& bK\cdot bL+bK\cdot b^*K-bL\cdot b^*L+b^*L\cdot b^*K,\\
B&\equiv& bT-b^*T,
\end{eqnarray*}
(temporary we omit superscript 1 for the modes of the first string).

We put (\ref{bb}) to (\ref{spusk2}) and isolate the exponent with
modes of the second and third strings as $\exp(-W_2)$. So we get
\begin{equation*}
{}_{32}\langle+|e^{-W_2}{}_1\langle+|e^{-W_1-b_0^1w}(A+b_0B)e^{b^*Sc^*}|+\rangle_1.
\end{equation*}
Here $w$, $W_1$ and $W_2$ have the same form as in the previous
section. Let us expand the exponent in $b_0^1$ mode (here we drop
the factor ${}_{32}\langle +|e^{-W_2}$)
\begin{eqnarray*}
&&{}_{1}\langle+|e^{-W_1}(1-b_0w)(A+b_0B)e^{b^*Sc^*}|+\rangle_1\\
&&={}_{1}\langle+|e^{-W_1}(A-b_0w A+b_0B)e^{b^*Sc^*}|+\rangle_1.
\end{eqnarray*}
The first term does not contribute. In two remaining terms we move
$b_0$ to the left vacuum:
\begin{equation}
\label{www} ={}_1\langle-|e^{-W_1}Be^{b^*Sc^*}|+\rangle_1-
{}_1\langle-|e^{-W_1}Aw e^{b^*Sc^*}|+\rangle_1.
\end{equation}
The direct calculation for (\ref{www}) leads to:
\begin{eqnarray*}
&&\langle V_3^{gh}|V_1^{gh}\rangle=-2\det(Z^{-1})
{}_{23}\langle+|\exp\left\{-\lambda^bZS\lambda^c-\sum_{n=0,\atop
m=1}^\infty\sum_{p,q=2}^3b_n^pX_{nm}^{pq}c_m^q\right\}\\
&&\cdot\left[\lambda^bZT+2\bar XZL\cdot \lambda^bZK-2\bar XZK\cdot
\lambda^bZL -2\lambda^bZL\cdot \lambda^bZK\cdot\bar U^{1q}c^q\right
].
\end{eqnarray*}
In the exponent we got the matrices $U^{pq}_{nm}$ (see eq.
(\ref{U})) as in the previous paragraph. This is not a surprise
because the pre-exponential factor does not contribute to the
exponent. For the following calculations we use the property $\bar
XZL=\bar XZK$ proved numerically (this property is true for any
level). So the result is
\begin{equation}
=-2\det(Z^{-1}){}_{23}\langle+|e^{-\check{b}^pU^{pq}c^q}[\theta
\check{b}^rX^{r1}ZT +2\check{b}^pX^{p1}ZL\cdot
\check{b}^rX^{r1}ZK\bar U^{1q}c^q],
\end{equation}
where $\theta\equiv(1-2\bar XZL)$ and
$\check{b}^r\equiv(b_0^r,b_1^r,b_2^r,...)$.

It is instructive to stress that $\check{b}^pX^{p1}$ does contain
the zero modes as well as $\check{b}^pU^{pq}$ in the exponent.

As for the first method we analyze the structure of the
pre-exponential factor in $b_0^{2,3}$. The free from $b_0$ term has
the form
\begin{equation}
\theta b^p\hat X^{p1}ZT +2b^p\hat X^{p1}ZL\cdot b^r\hat X^{r1}ZK\bar
U^{1q}c^q.
\end{equation}
Due to the normalization on $\mathfrak{J}_0$ (see below) this term
vanishes. On the Figure \ref{free-terms-non}  we can see the
behavior of two free terms.
\begin{figure}[h!]
\begin{center}
\includegraphics[scale=0.8]{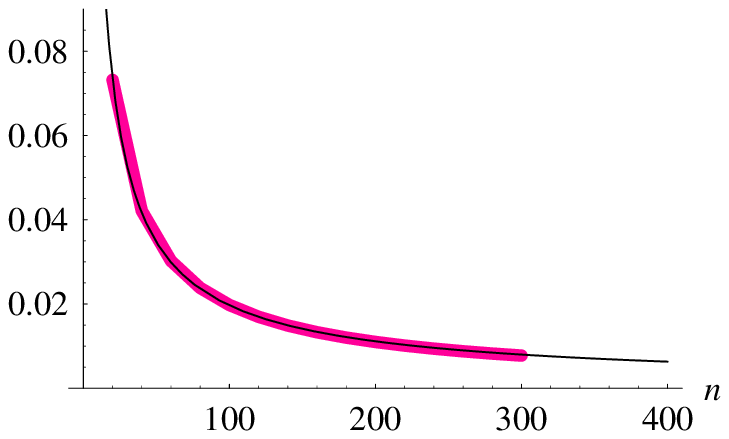}
\quad\includegraphics[scale=0.8]{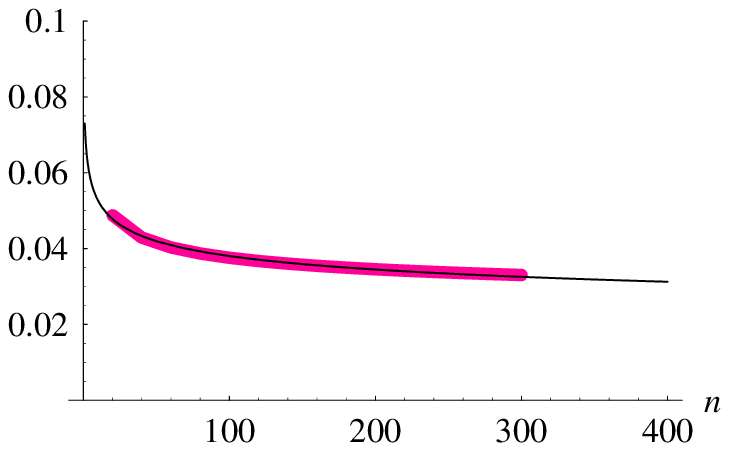} \caption{}
\label{free-terms-non}
\end{center}
\end{figure}
As we can see the both terms vanish as $\sim\frac{0.856}{n^{0.819}}$
and $\sim\frac{0.073}{n^{0.142}}$ respectively.

Let us consider the terms proportional to $b_0^2$ and $b_0^3$:
\begin{eqnarray*}
&&-\theta b^2_0\bar X^{21}ZT-\theta b^3_0\bar X^{31}ZT \\
&&-2(b^2_0\bar X^{21}ZL\cdot b^p\hat X^{p1}ZK+b^3_0\bar
X^{31}ZL\cdot b^p\hat X^{p1}ZK)\bar U^{1q}c^q\\
&&+2(b^2_0\bar X^{21}ZK\cdot b^p\hat X^{p1}ZL+b^3_0\bar
X^{31}ZK\cdot b^p\hat X^{p1}ZL)\bar U^{1q}c^q\\
&&\equiv b_0^2(\mathfrak{J}_0+b^2\mathfrak{\bar A}^2 \bar
U^{1q}c^q+b^3\mathfrak{\bar A}^3\bar U^{1q}c^q)+ b_0^3(\mathfrak{J}
_0'+b^2 \mathfrak{\bar B}^2\bar U^{1q}c^q+b^3\mathfrak{\bar B}^3\bar
U^{1q}c^q),
\end{eqnarray*}
where we define
\begin{eqnarray}
\mathfrak{J}_0&=&-\theta\bar X^{21}ZT,\qquad
\mathfrak{J}_0'=-\theta\bar
X^{31}ZT,\nonumber\\
\mathfrak{\bar A}^i&=&-2(\hat X^{i1}ZK\cdot\bar X^{21}ZL-\hat
X^{i1}ZL\cdot\bar
X^{21}ZK),\quad i=2,3,\\
\mathfrak{\bar B}^i&=&-2(\hat X^{i1}ZK\cdot\bar X^{31}ZL-\hat
X^{i1}ZL\cdot\bar X^{31}ZK).\nonumber
\end{eqnarray}
For the correct structure of vacua it's necessary but not sufficient
to fulfil the condition $\mathfrak{J}_0'+\mathfrak{J}_0\to0$. By
numerically calculations: $\mathfrak{J}_0'+\mathfrak{J}_0=0$,
starting from the first level.

In the pre-exponential factor we have the term which is quadratic in
$b_0$. It has the form
\begin{equation}
2b^2_0b^3_0(-\bar X^{21}ZL\cdot\bar X^{31}ZK+\bar X^{31}ZL\cdot\bar
X^{21}ZK)\bar U^{1q}c^q.
\end{equation}

The next step is to rise up the linear and quadratic terms to the
exponent as we did in the previous section. In terms of that method
we again get some new matrices $V^{pq},~p=2,3$. So we have the
conditions (\ref{**}) and (\ref{conditions}) and the following
system of equations:
\begin{subequations}
\label{beta}
\begin{eqnarray}
&1.&\mathfrak{J}_0(\bar V^{2q}+\bar V^{3q})=2(\bar X^{21}ZL\cdot\bar
X^{31}ZK-
    \bar X^{31}ZL\cdot\bar X^{21}ZK)\bar U^{1q},\label{beta-1}\\
&2.&(\bar V^{2q}+\bar V^{3q})c^q\cdot b^p\hat V^{ps}c^s=0,\label{beta-2}\\
&3.&\hat V^{pq}=\frac{\mathfrak{\bar
A}^p}{\mathfrak{J}_0}\otimes\bar U^{1q},\label{beta-3}\\
&4.&\hat V^{pq}=\frac{\mathfrak{\bar
B}^p}{\mathfrak{J}_0}\otimes\bar U^{1q}.\label{beta-4}
\end{eqnarray}
\end{subequations}
We get the same structure as in eqs. (\ref{alpha}). However here we
have the equations in terms of the matrices $X$ versus matrices $U$
as in the previous section.

\begin{figure}[h]
\begin{center}
\includegraphics[scale=0.8]{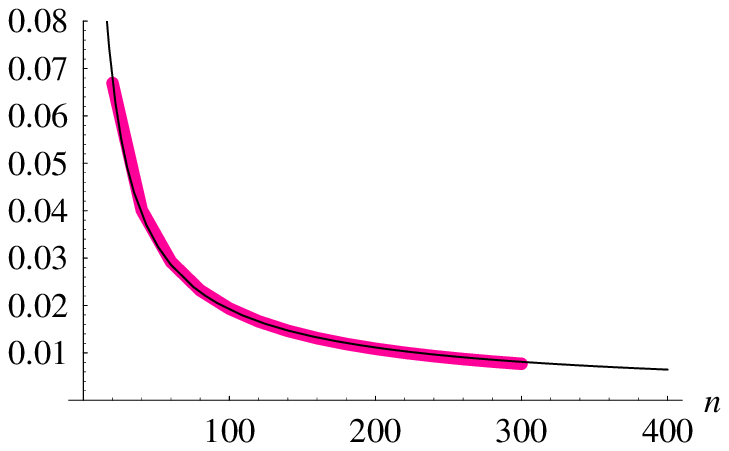}
\caption{} \label{consist-non}
\end{center}
\end{figure}

The consistency conditions of the system is

$$
\frac{\mathfrak{\bar A}^r}{\mathfrak{J}_0}+\frac{\mathfrak{\bar
B}^r}{\mathfrak{J}_0}\to0,\quad r=2,3.
$$
The numerical calculations are summarized in Figure
\ref{consist-non}. Data produce the fit
$\sim\frac{0.705}{n^{0.783}}$, i.e. the data seem to indicate that
the sum vanishes as $n\to\infty$.

To solve the system (\ref{beta}) we use the same strategy as in the
previous section.

\begin{figure}[h!]
\begin{center}
\includegraphics[scale=0.8]{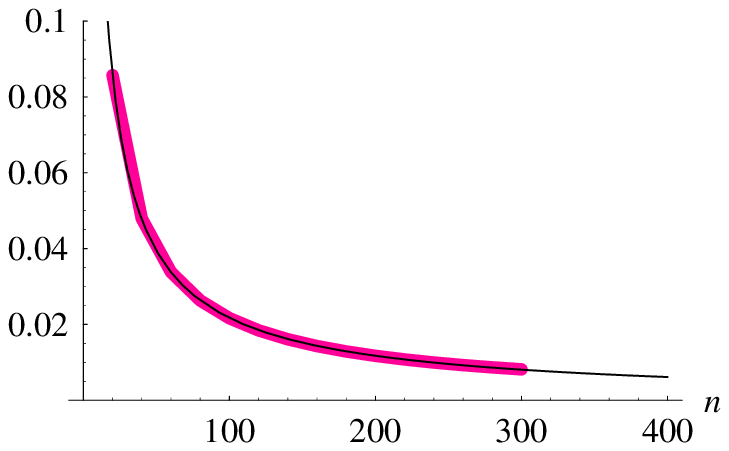}
\caption{} \label{UU22+VV22-non}
\end{center}
\end{figure}

Let us stress that the system (\ref{beta}) defines only the sum of
the rows $\bar V^{2q}+\bar V^{3q}$. We set $\bar U^{pq}=-\bar
V^{pq}$ thus the equation (\ref{**}) is fulfilled. So the equation
\eqref{beta-1} in the system transforms into
$$
\mathfrak{J}_0(\bar U^{2q}+\bar U^{3q})=-2(\bar X^{21}ZL\cdot\bar
X^{31}ZK-
    \bar X^{31}ZL\cdot\bar X^{21}ZK)\bar U^{1s}.
$$
Using $\bar U^{1q}=-(\bar U^{2q}+\bar U^{3q})$, rewrite it as
$$
\mathfrak{J}_0=2(\bar X^{21}ZL\cdot\bar X^{31}ZK-
    \bar X^{31}ZL\cdot\bar X^{21}ZK)
$$
or
$$
-(1-2\bar XZL)\bar X^{21}Z(L-K)=2(\bar X^{21}ZL\cdot\bar X^{31}ZK-
    \bar X^{31}ZL\cdot\bar X^{21}ZK).
$$
On the Figure \ref{UU22+VV22-non} we present the behavior of this
equation rewritten in the form
$$
1-2\frac{\bar X^{21}ZL\cdot\bar X^{31}ZK-
    \bar X^{31}ZL\cdot\bar X^{21}ZK}{\mathfrak{J}_0}
$$
for $n\to\infty$. For this graph we have the fit $\sim
\frac{1.13}{n^{0.86}}$.

Eq. \eqref{beta-3} in the system defines $\hV^{pq}$ and we can check
the conditions (\ref{conditions}). The numerical calculations shows
these conditions to behave similarly. So we give only one result for
the first expression $Max|\hat U^{22}+\hat V^{22}|$ for
$n=40,...,300$. We discard the last 20 rows and columns in the
matrices in our calculations.

\begin{figure}[h!]
\begin{center}
\includegraphics[scale=0.8]{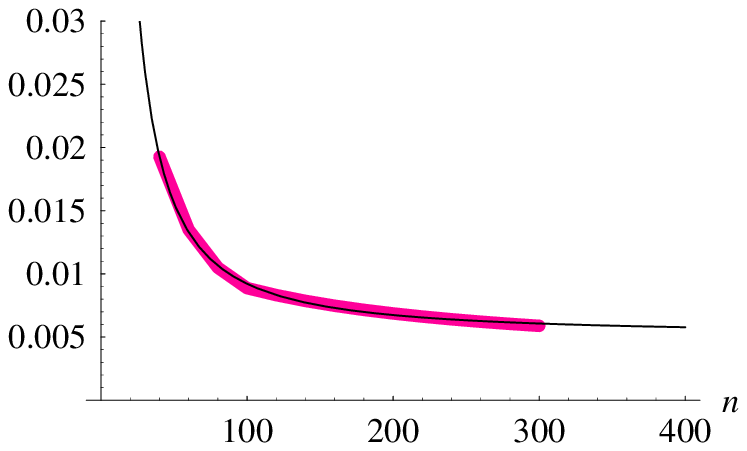}
\caption{} \label{U22+V22-non}
\end{center}
\end{figure}

For $Max|\hat U^{22}+\hat V^{22}|$ we got the fit of the form $\sim
0.005+\frac{2.17}{n^{1.36}}$.

Hence we got that the descent relation in the ghost sector has the
following form
\begin{eqnarray}
{}_{321}\langle V_3^{gh}|V_1^{gh}\rangle_1&=&\mathfrak{J}_0 N_3\frac
i 2\det(1-SX){}_{23}\langle +|
(b_0^2-b_0^3)e^{-b^2Sc^3-b^3Sc^2}\nonumber\\
&=& \mathfrak{J}_0 N_3\frac i 2\det(1-SX){}_{23}\langle V_2^{gh}|
\equiv \tilde{Z}_{nov}^{gh}{}_{23}\langle V_2^{gh}|.
\end{eqnarray}
Here we restore the coefficients which were dropped above.

\subsubsection{Result with Matter Sector}
The normalization $\tilde{Z}_3$ in the descent relation $\langle
V_3|V_1\rangle$ consists of two factors: ghost $\tilde{Z}^{gh}$ and
matter $\tilde{Z}^m$ ones, i.e. $\tilde{Z}_3=\tilde{Z}^m
\tilde{Z}^{gh}$. In the ghost sector in the ``overlap calculations''
the normalization is
\begin{equation}
\tilde{Z}_{ov}^{gh}=-\frac i 4 N_3 J_0 \det(1-SX),
\end{equation}
in the ``non-overlap calculations'' we got
\begin{equation}
\tilde{Z}_{nov}^{gh}=\frac i 2 N_3 {\mathfrak J}_0 \det(1-SX).
\end{equation}
In the matter sector we have the following normalization factor
\begin{equation}
\label{gamma-matter} \tilde{Z}^m=\det(1+V^{11m}S)^{-13}.
\end{equation}
 In (\ref{gamma-matter}) the matrix $V^{11m}$ is the Neumann
matrix for matter \cite{GJ,T}. This result was given in \cite{KF}
and we checked its accuracy. Because of the simple structure of the
vertices in the matter sector one can directly use \cite{KP} without
any intermediate calculations.

\begin{figure}[h]
\begin{center}
\includegraphics[scale=0.8]{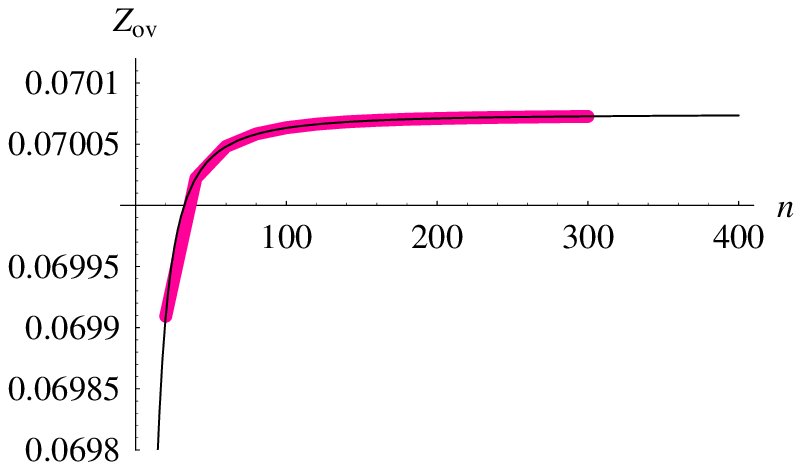}\qquad
\includegraphics[scale=0.8]{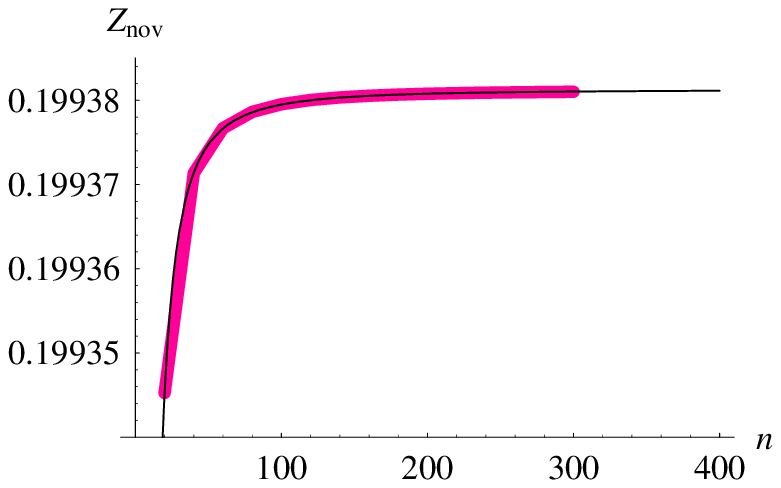}
\caption{} \label{graphs-for-gammas}
\end{center}
\end{figure}

Thus the numerical calculations of the full normalization factors
give us that we have two different results which are displayed on
the Figure \ref{graphs-for-gammas}.

We have the following fits for these graphs: for the first,
``overlap calculations'' we have $\tilde{Z}_3\sim
0.070075-\frac{0.024}{n^{1.666}}$; for the second, ``non-overlap
calculations'' we have $\tilde{Z}_3\sim
0.199381-\frac{0.01}{n^{1.87}}$.

To conclude we calculate the descent relation for $\langle
V_3|V_1\rangle$ using two different schemes. Both calculations are
performed with using the oscillator level truncation method and
differ in a treatment of the mid-point insertion. Both schemes gives
the vertex $\langle V_2|$ with two different normalization factors.
A similar problem with the oscillator level truncation method was
been found in \cite{Okuyama:2002tw}.

\section*{Acknowledgements}

We would like to thank D.~Belov and A.~Koshelev for useful
discussions and E. Fuchs and M. Kroyter for the correspondence and
usefull remarks. The work is supported in part by RFBR grant
05-01-00758 and Russian President's grant NSh-2052.2003.1. The work
of I.A. is supported in part by INTAS grant 03-51-6346.
\section*{Appendix}

In this article we use the notations for the vacua following
\cite{GJ}

 in the matter sector:
\begin{equation*}
a_n|0\rangle=0,\qquad n\geq0,\qquad \langle 0|0\rangle=1,
\end{equation*}

in the ghost sector:
\begin{eqnarray*}
c_n|+\rangle&=&0, \quad n\geq0,\\
b_m|+\rangle&=&0, \quad m\geq1,
\end{eqnarray*}
$$
|-\rangle=b_0|+\rangle,\qquad \langle -|+\rangle=\langle
+|-\rangle=1.
$$
The vertex $\langle V_3|$ is given by:
$$
\langle
V_3|={}_{321}\langle+|\exp\left\{-\sum_{r,s=1}^3\sum_{n=0,\atop
m=1}b_n^rX_{nm}^{rs}c_m^s\right\} N_3,
$$
where
$$
{}_{321}\langle +|\equiv{}_3\langle +|{}_2\langle +|{}_1\langle +|.
$$
 Neumann matrix obey the following properties  \cite{GJ,KP}
\begin{eqnarray*}
X_{nm}^{r,r}+X_{nm}^{r,r+1}+X_{nm}^{r,r-1}&=&S_{nm} ,\quad
n\neq0\\
X_{0m}^{r,r}+X_{0m}^{r,r+1}+X_{0m}^{r,r-1}&=&0.\nonumber
\end{eqnarray*}
here $X_{nm}^{r,r+2}=X_{nm}^{r,r-1}$ we have the same formula for
the second fixed index. We separate the zero mode in the insertion
$b_+b_-$ as
$$
b_+\equiv b(\pi/2)=\sum_{-\infty}^\infty i^nb_n=bK+b_0+b^*L,\quad
b_-\equiv b(-\pi/2)=\sum_{-\infty}^\infty i^{-n}b_n=bL+b_0+b^*K,
$$
where
\begin{eqnarray*}
b^j\equiv(b_1^j,b_2^j,...),\quad
b^{j*}\equiv(b_{-1}^j,b_{-2}^j,...), \quad
c^{j*}\equiv(c^j_{-1},c^j_{-2},...)^T,\quad
c^j\equiv(c^j_1,c^j_2,...)^T,\\
K\equiv(i,-1,-i,1,...)^T,\quad L\equiv(-i,-1,i,1,...)^T,\quad
T\equiv L-K=(-2i,0,2i,0,...)^T
\end{eqnarray*}
In the vertex $V_3$ we separate the dependence on the first string
as follows
\begin{eqnarray*}
W_1&=&\sum_{n,m=1}^\infty
b^1_nX_{nm}^{11}c_m^1+\sum_{s=2}^3\sum_{n,m=1}^\infty
b_n^1X^{1s}_{nm}c_m^s+\sum_{r=2}^3\sum_{n=0,\atop m=1}^\infty
b_n^rX^{r1}_{nm}c_m^1\\
&\equiv& b^1\hat Xc^1+b^1\lambda^c+\lambda^bc^1.
\end{eqnarray*}
where
\begin{equation*}
\lambda^c\equiv\sum_{s=2}^3\hat X^{1s}c^s,\quad
\lambda^b\equiv\sum_{r=2}^3(b_0^r\bar X^{r1}+b^r\hat X^{r1}),
\end{equation*}
and
\begin{equation*}
W_2=\sum_{n,m=1}^\infty b^p_nX_{nm}^{11}c_m^q,\qquad p,q=2,3.
\end{equation*}


\end{document}